\documentclass[twocolumn,aps,superscriptaddress,floatfix,final,showpacs,prb,eqsecnum]{revtex4}
\usepackage{amsmath}
\usepackage{graphicx}
\usepackage{amssymb}

\makeatletter
\usepackage{epsfig}
\usepackage{dcolumn}
\usepackage{bm}
\usepackage{amsfonts}
\usepackage{graphicx}

\begin{document}

\title{Influence of disorder on incoherent transport near the Mott transition}

\author{Milo\v s M. Radonji\'{c}}
\affiliation{Scientific Computing Laboratory, Institute of Physics
Belgrade, Pregrevica 118, 11080 Belgrade, Serbia}
\author{D. Tanaskovi\'{c}}
\affiliation{Scientific Computing Laboratory, Institute of Physics
Belgrade, Pregrevica 118, 11080 Belgrade, Serbia}
\author{V. Dobrosavljevi\'{c}}
\affiliation{Department of Physics and National High Magnetic
Field Laboratory, Florida State University, Tallahassee, Florida
32306, USA}
\author{K. Haule}
\affiliation{Department of Physics and Astronomy, Rutgers
University, Piscataway, New Jersey 08854, USA}

\begin{abstract}
We calculate the optical and DC conductivity for half-filled
disordered Hubbard model near the Mott metal-insulator transition.
As in the clean case, large metallic resistivity is driven by a
strong inelastic scattering, and Drude-like peak in the optical
conductivity persists even at temperatures when the resistivity is
well beyond the semiclassical Mott-Ioffe-Regel limit. Local random
potential does not introduce new charge carriers, but it induces
effective local carrier doping and broadens the bandwidth. This
makes the system more metallic, in agreement with the recent
experiments on X-ray irradiated charge-transfer salts.

\end{abstract}


\maketitle

\section {Introduction}

Thermodynamic and transport properties of many strongly correlated
materials are influenced by inhomogeneities and
disorder,\cite{Miranda2005} and their importance becomes even more
evident with the advances in experimental
techniques.\cite{Basov2007,Davis2007} Very recently, the effects
of disorder on the optical and DC conductivity of the organic
charge-transfer salts have been systematically explored by
introducing defects by X-ray irradiation.\cite{Analytis2006,
Sasaki2007, Sasaki2008} The conductivity has proven to be very
sensitive on the duration of the irradiation, and different
physical mechanisms were advocated to explain such a
behavior.\cite{Analytis2006, Sasaki2007, Sasaki2008}

Members of $\kappa$-family of organic charge-transfer salts have
half-filled conduction bands with the effective Coulomb repulsion
comparable to the bandwidth,\cite{Powell2006} and their proximity
to the Mott metal-insulator transition can be tuned by doping or
applying the pressure. On the metallic side of the Mott
transition, the  Fermi liquid transport at low temperatures is
followed by an incoherent transport  at higher temperatures
dominated by the large scattering rate, and  with resistivities an
order of magnitude larger than the Mott-Ioffe-Regel (MIR)
limit,\cite{Gunnarsson2003,CalandraPRL2001,CalandraPRB2002,Allen2002,Hussey2004}
which is the maximal resistivity that can be reached in a metal
according to the Boltzmann semiclassical theory. From the
theoretical point of view, the violation of the MIR condition and
the appearance of the maximum in the resistivity temperature
dependence is not easy to explain. However, at least for
$\kappa$-organics, a significant progress has been recently
achieved when the transport properties were successfully described
even on the quantitative level within the dynamical mean field
theory (DMFT).\cite{Merino2000,Limelette2003,Merino2008}
Therefore, it is important to explore the effects of X-ray
irradiation within the same theoretical framework by introducing
the disorder into the model.

Most of the theoretical work on the influence of disorder on the
physical properties near the Mott transition have been so far
restricted to binary disorder distribution,\cite{Laad2001} or low
temperatures where the DMFT has been extended in order to
incorporate the Anderson localization effects.\cite{
Dobrosavljevic1997,Dobrosavljevic2003,Byczuk2005} However, for a
comparison with the experiments on $\kappa$-organics, we need the
results in a wide temperature range, and since the disorder is
gradually generated by X-ray irradiation, the simplest approach of
disorder averaging on the level of coherent-potential
approximation (CPA),\cite{Economou2005} that we apply in our work,
should be sufficient to explain the main modifications in the
optical and DC conductivity caused by the disorder. We find that,
as in the clean case, it is the large electron-electron scattering
that leads to the destruction of coherent quasiparticles and to
the resistivity well beyond the MIR limit on the metallic side of
the Mott transition. For a fixed interaction strength, however,
the random potential effectively weakens the correlation effects
and moves the system away from the Mott transition. Therefore, the
disorder increases the conductivity in agreement with the
experiments on the organic charge-transfer salts.

The rest of the paper is organized as follows. Section II contains
the DMFT equations for the disordered Hubbard model, brief
discussion on the impurity solvers that we use and comments on the
validity of the DMFT approximation. Section III presents the
results for the temperature dependence of the density of states,
optical conductivity and DC resistivity near the Mott transition
for the pure and disordered system. Our results are compared with
the experiments on X-ray irradiated $\kappa$-organics in Section
IV. Section V contains the conclusions.

\section {Disordered Hubbard model}

We consider half-filled Hubbard model with site-diagonal disorder
as given by the Hamiltonian
\begin{equation} H=-\sum_{ij,\sigma}t_{i,j}c^{\dag}_{i\sigma}c_{j\sigma}+U\sum_{i}n_{i\uparrow}n_{i\downarrow}+\sum_{i\sigma}v_{i}n_{i\sigma} -\mu \sum_{i\sigma}n_{i\sigma} \label{eq:habard.ham}.
\end{equation}
Here $t_{i,j}$ is the hopping amplitude, $U$ the interaction
strength, $c_{i \sigma}^{\dag}$ is the creation operator, and
$n_{i \sigma}=c_{i \sigma}^{\dag}c_{i \sigma}$ the occupation
number operator on site $i$ and for spin $\sigma$. Half-filling
condition is enforced by the chemical potential $\mu$. We model
the disorder by random energies $v_i$ taken from the uniform
distribution in the interval $(-W/2,W/2)$.

In the dynamical mean field theory, which is formally exact in the
limit of large coordination number, the Hubbard model reduces to a
model of an Anderson impurity in a conduction bath which has to be
determined self-consistently. In the presence of disorder, we need
to consider an ensemble of impurities, and the conduction bath is
determined in the process of averaging over the disorder.

The central quantity in DMFT is the local Green function,
$G_{i\sigma}(\tau-\tau')=-\langle
Tc_{i\sigma}(\tau)c_{i\sigma}^{\dag}(\tau') \rangle_{S_{eff}^i}$,
which is a site-dependent quantity in the presence of disorder.
The local effective action is given by
\begin{eqnarray}
&&S_{eff}^i=-\frac{1}{\beta} \sum_{i
\omega_n,\sigma}c^{\dag}_{i\sigma}(i\omega_n)
[i\omega_{n}+\mu-v_{i}-\Delta(i\omega_{n})]c_{i\sigma}(i \omega_n)
\nonumber \\  &&+\frac{1}{\beta}U\sum_{i \omega_n} \,
{n_{\uparrow}(i \omega_n)n_{\downarrow}(i \omega_n)},
\label{eq:action}
\end{eqnarray}
where $\Delta$ is the conduction bath whose self-consistent value
will be obtained  in the iterative procedure. The quantity that we
average over the disorder is the local Green function,
$G_{av}(i\omega_n)=\int dv P(v) G(i\omega_n, v)$. Though we
consider a continuous distribution of disorder, $P(v)$, in
practice it is sufficient to take a finite number of random
energies, and the integral replaces with a sum. In the case of
uniform disorder
\begin{equation}
G_{av} (i\omega_{n})=\frac{1}{N}\sum_{i=1}^NG_i(i\omega_n).
\label{eq:gave}
\end{equation}

The averaged Green function $G_{av}$ and the conduction bath
$\Delta$ determine the self-energy through the relation
\begin{equation}
G^{-1}_{av}(i\omega_n)=i\omega_n+\mu-\Delta(i\omega_n)-\Sigma(i\omega_n).
\label{eq:save}
\end{equation}
The self-consistency condition follows from the assumption that
the lattice self-energy coincides with the impurity self-energy.
Then the disorder averaged local Green function has to be equal to
the local component of the lattice Green function,
\begin{equation}
G_{av}(i\omega_{n})=\int
d\varepsilon{\frac{D(\varepsilon)}{i\omega_n+\mu-\varepsilon-\Sigma(i\omega_n)}}.
\label{eq:loop}
\end{equation}
Here $D(\varepsilon)$ is the density of states in the absence of
disorder and interaction. Eq.~(\ref{eq:save}) determines new
conduction bath which completes the self-consistency loop. We note
that our treatment of disorder reduces to the CPA approximation in
the absence of interaction.

The most difficult step in the solution of DMFT equation is the
calculation of the local Green function from the Anderson impurity
action Eq.~(\ref{eq:action}). In this paper, we solve the Anderson
impurity model with One-Crossing Approximation
(OCA),\cite{Pruschke1993,Haule2001} and cross-check the results
with Continuous Time Quantum Monte Carlo (CTQMC) impurity
solver.\cite{Werner2006,Haule2007} The OCA impurity solver has an
advantage that it gives a solution on real frequency axis, which
is necessary for a calculation of the response functions. Except
at the lowest temperatures, where the OCA impurity solver does not
reproduce the Fermi-liquid behavior, the results obtained with
these two impurity solvers are qualitatively the same  and
quantitatively very similar.

Since we are going to compare the results with the experiments on
$\kappa$-organics, we briefly comment on the validity of DMFT
approximation. The main advantage of this method is that it treats
on an equal footing low and high energy part of the spectrum and
fully takes into account the inelastic electron-electron
scattering. Therefore, it successfully describes a crossover from
the low temperature Fermi liquid regime to the incoherent
transport at higher temperatures, which is typical for many
strongly correlated systems. The DMFT is based on the assumption
of locality of the self-energy, and hence it does not fully take
into account intersite correlations. However, the
$\kappa$-organics have weakly anisotropic triangular lattice
structure\cite{Powell2006} which makes the intersite correlations
smaller due to the geometrical frustration. In this case, local
DMFT approximation gives similar results as generalized cluster
DMFT.\cite{Liebsch2009} Finally, since the lattice structure
enters the DMFT equations only through the density of states, the
transport properties does not depend much on the details of the
band structure, and we will consider the hypercubic lattice which
has the density of states in the form of a Gaussian $
D(\varepsilon)= \sqrt{ \frac{2}{\pi}} \, \, e^{-2 \varepsilon^2}$,
where the energy is given in units of the half-bandwidth.

\section{Conductivity near the Mott transition}

The phase diagram of half-filled Hubbard model in DMFT
approximation is well known.\cite{Georges1996} At low
temperatures, an increase of the interaction $U$ leads to the
first order metal-insulator transition. The line of the first
order phase transition ends in the critical point at temperature
$T_c$ and interaction $U_c$. In this work we will concentrate on
the values of interaction equal and slightly lower than $U_c$. At
these values of $U$, there is a gradual crossover from the Fermi
liquid to an incoherent metallic and, eventually, insulating
behavior as the temperature is increased.

\begin{figure*}[t]
\begin{center}
\includegraphics[  width=6 in,
keepaspectratio]{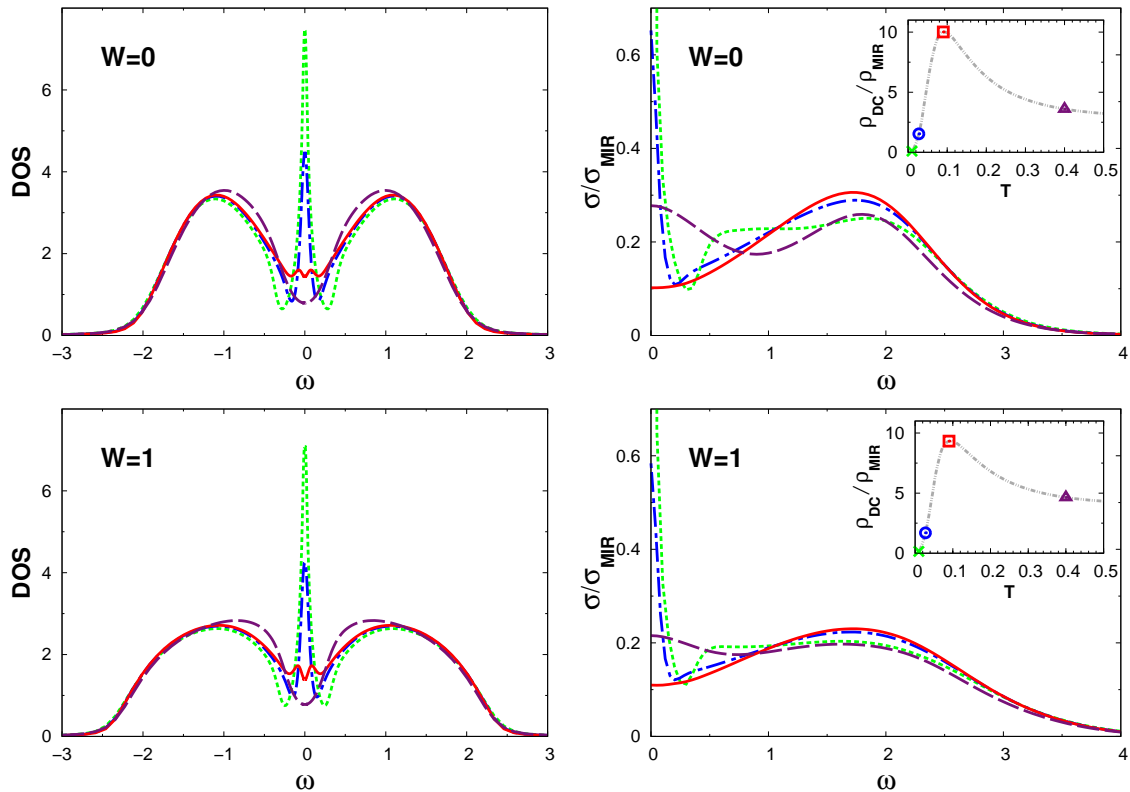}
\caption{(Color online) Density of states and optical conductivity
as a function of frequency in the clean case for $U=0.94 \,
U_c|_{W=0}$ (upper panel) and disordered case, $U=0.94 \,
U_c|_{W=1}$ (lower panel). Different colors correspond to the four
distinctive transport regimes (see the text). The insets show the
temperature dependence of DC resistivity. $T$, $\omega$ and $W$
are given in units of bare $E_F$.}
\label{conductivity}%
\end{center}
\end{figure*}

The central quantity that we calculate is the optical
conductivity. In DMFT it is given by\cite{Georges1996}
\begin{eqnarray}
&&\sigma(\omega)=\frac{\pi e^2}{\hbar }
\int_{-\infty}^{+\infty}d\epsilon \int_{-\infty}^{+\infty}d\nu
D(\epsilon)\rho(\epsilon,\nu)\rho(\epsilon,\nu+\omega) \nonumber
\\ &&\times \frac{f(\nu)-f(\nu+\omega)}{\omega}.
\end{eqnarray}
Here $\rho(\epsilon,\nu)=-\frac{1}{\pi}\mbox{Im}G(\epsilon,\nu)$,
and $G(\epsilon,\nu)=(\nu + \mu -\epsilon -\Sigma(\nu))^{-1}$. We
will express the conductivity in units of the Mott-Ioffe-Regel
limit for minimal metallic conductivity. The MIR limit,
$\sigma_{_{MIR}}$, is the conductivity which is reached when the
electron mean free path becomes comparable to the lattice spacing,
$l \sim a$. According to the semiclassical arguments, the
electrons can scatter at most on every atom and the conductivity
in a metal cannot be smaller than $\sigma_{_{MIR}}$. For
half-filled hypercubic lattice (which has Gaussian density of
states), the MIR condition $l = a$ is equivalent to $E_{_F} \tau
=1$, where $E_{_F}$ is the bare Fermi energy, i.e. ~half-bandwidth
of the noninteracting electrons, and $\tau^{-1}$ is the scattering
rate. Here $\hbar$ is set to 1. Therefore, the MIR limit is set by
a condition
\begin{equation}
 \tau_{_{MIR}}^{-1} =-2 \mbox{Im} \Sigma(0^+)=1,
\end{equation}
where $\Sigma$ is the self-energy measured in units of $E_F$.

The density of states and optical conductivity for a clean system
and in a presence of moderate disorder, $W=1$, are shown in
Fig.~\ref{conductivity}. The disorder effectively increases the
bandwidth and the critical interaction $U_c$. In our case, we find
that $U_c|_{W=0}=2.2$ and $U_c|_{W=1}=2.45$. The increase of $U_c$
due to disorder is in agreement with earlier estimates obtained by
iterated perturbation theory.\cite{Aguiar2005} The critical
temperature $T_c$ weakly depends on the disorder strength,
$T_c|_{W=1} \approx T_c|_{W=0}=0.04$, where $k_B$ is set to 1. On
Fig.~\ref{conductivity} we compare the data at the same relative
value $U/U_c=0.94$, and for several characteristic temperatures.
We see that the disorder does not lead to qualitative differences
and if the interaction is the same when scaled with $U_c$, the
density of states and the optical conductivity are even
quantitatively very similar.

\begin{figure}[t]
\begin{center}
\includegraphics[  width=2.5 in,
keepaspectratio]{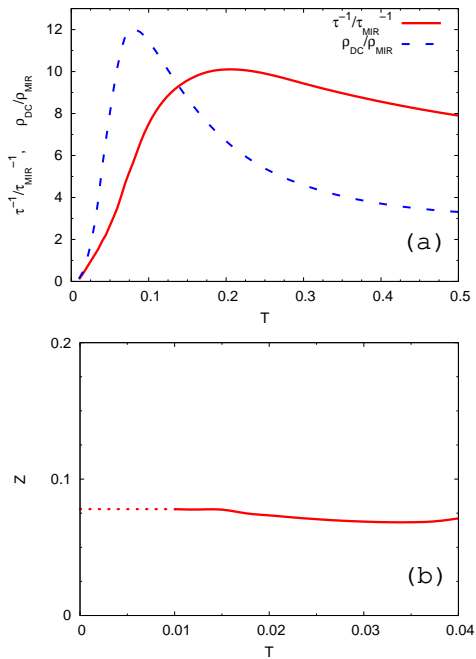} \caption{(Color online) (a)
Scattering rate (full line) and DC resistivity (dashed) as a
function of temperature. (b) Quasiparticle weight as a function of
temperature. The data are for the clean system at $U=0.95 \,
U_c$.}
\label{inelastic}%
\end{center}
\end{figure}

We can identify several regimes of the electron transport. At low
temperature (green dotted lines and crosses in the insets) the
scattering rate, $\tau^{-1}=-2 \mbox{Im} \Sigma(0^+)$, is small
and the transport is dominated by long-lived coherently
propagating quasiparticles. The blue dash-dotted lines (blue
circles in the insets) correspond to the temperature when the
resistivity is already larger then the MIR limit, the scattering
rate $\tau^{-1}$ is larger than $E_F$, and the Fermi liquid
picture of well-defined quasiparticles ceases to be valid.
However, a Drude-like peak in the optical conductivity, as well as
a peak in the density of states, are still present. Our results
show that this is the case also in the presence of moderate
disorder. The resistivity maximum (red full line and square) is
reached when the peak at the Fermi level is fully suppressed and
when a dip at the Fermi level appears both in the density of
states and in the optical conductivity. The resistivity maximum is
more than an order of magnitude larger than
$\rho_{_{MIR}}=\sigma_{_{MIR}}^{-1}$. At even higher temperatures
(violet dashed line and triangle) low frequency optical
conductivity increases due to the thermal excitations.

Fig.~\ref{inelastic} helps us to further distinguish the mechanism
leading to the large resistivity and its strong temperature
dependence. We see that the scattering rate gives the main
contribution to the resistivity temperature dependence and causes
the violation of the MIR limit, Fig.~\ref{inelastic}(a), while the
quasiparticle (Drude) weight $Z=(1+|\partial \mbox{Re}\, \Sigma
(\omega) / \partial \omega |_{\omega=0})^{-1}$ is almost
temperature independent, Fig.~\ref{inelastic}(b). The dotted part
of the line is an extrapolation of the OCA results to zero
temperature. We have also checked that $Z$ depends very weakly on
the temperature using numerically exact CTQMC impurity solver.
Therefore, we can conclude that the driving mechanism for large
resistivity is the large scattering rate and not the reduction of
the spectral weight near the Fermi level. This feature, already
seen in the experiments on $\mbox{VO}_2$\cite{Basov2006} and
charge-transfer salts,\cite{Merino2008} seem to be common for the
systems with half-filled conduction band near the Mott transition.
This should be contrasted with the doped Mott insulators where the
main reason for the violation of the MIR condition is a decimation
of the Drude peak in the optical conductivity by the time MIR
limit is reached, which can be interpreted as a reduction of the
number of charge carriers.\cite{Gunnarsson2003,Hussey2004}

\begin{figure}[t]
\begin{center}
\includegraphics[  width=2.5 in,
keepaspectratio]{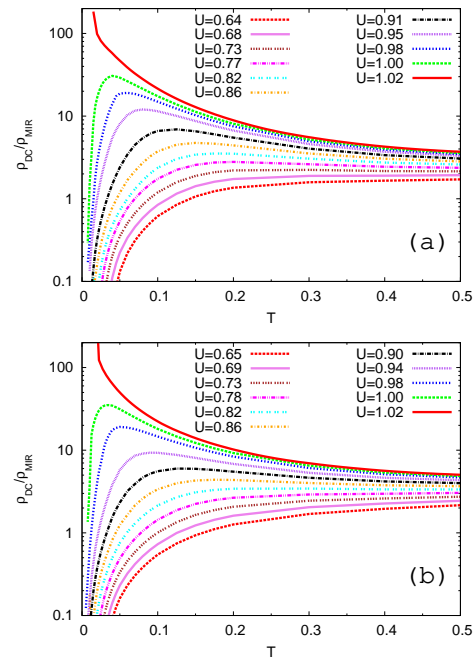}
\caption{(Color online) Temperature dependence of DC resistivity
for different interaction $U$ in the clean case, $W=0$ (a) and
disordered case, $W=1$ (b). U is given in units of $U_c(W)$.}
\label{resistivity}%
\end{center}
\end{figure}

The results for temperature dependence of DC resistivity,
$\rho_{_{DC}}=\sigma^{-1}(\omega \rightarrow 0)$, for several
values of interaction $U$ are shown in Fig.~\ref{resistivity}. The
resistivity is given in units $\rho_{_{MIR}}$. For clarity it is
shown on a logarithmic scale. The resistivity in the clean and
disordered case are even quantitatively very similar when the
interaction is scaled with $U_c(W)$.

\section{Increase of metallicity by disorder}

Very recent experiments\cite{Analytis2006, Sasaki2007, Sasaki2008}
on the charge-transfer organic salts provide a rather unique
opportunity to study the effects of disorder on transport
properties without changing external parameters or chemical
composition. The level of defects (disorder) directly depends on
the time of exposure to the X-rays. The optical and DC
conductivity are proven to be very sensitive on irradiation time
showing an increase in the  conductivity with the time of
irradiation. The experiments measured both interlayer and in-plane
resistivity with similar conclusions. Different physical
mechanisms were proposed to explain the increase of conductivity.
Analytis {\it et al.}\cite{Analytis2006} proposed a
defect-assisted interlayer conduction channel for the reduction of
resistivity, and Sasaki {\it et al.}\cite{Sasaki2007, Sasaki2008}
proposed that the irradiation leads to the effective doping of
carriers into the half-filled Mott insulator.

\begin{figure}[t]
\begin{center}
\includegraphics[  width=2.6 in,
keepaspectratio]{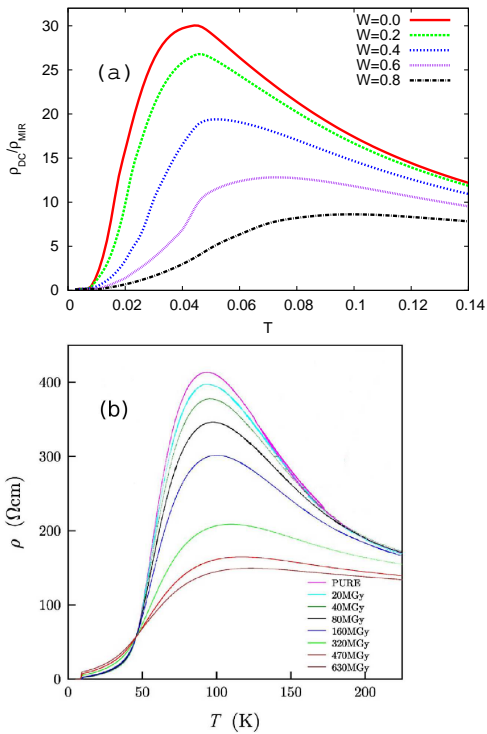}
\caption{(Color online) (a) Temperature dependence of DC
resistivity for fixed $U=2.2=U_c|_{W=0}$, and various levels of
disorder. (b) Experiments on $\kappa \mbox{-(BEDT-TTF)}_2
\mbox{Cu(SCN)}_2$, taken from Ref.~\onlinecite{Analytis2006}.}
\label{comparison}%
\end{center}
\end{figure}

The DMFT has successfully described the transport properties of
organic salts even on the quantitative
level.\cite{Limelette2003,Merino2008} In order to make a
comparison with the experiments with irradiation induced defects,
we solve the DMFT equations for fixed interaction $U$ and vary the
level of disorder $W$. The results for DC resistivity are shown in
Fig.~\ref{comparison}(a). The data for $T<0.01$ are obtained using
CTQMC impurity solver. The presence of even a weak disorder
significantly decreases the resistivity by effectively moving the
system away from the Mott insulator, as explained in the previous
section. Our data are very similar to the measurements on
charge-transfer salt $\kappa \mbox{-(BEDT-TTF)}_2
\mbox{Cu(SCN)}_2$ from Ref.~\onlinecite{Analytis2006}, which are
shown in Fig.~\ref{comparison}(b). We note that these data are for
interlayer resistivity while our DMFT calculation corresponds to
in-plane transport. However, the interlayer transport is due to
incoherent tunneling which is proportional to in-plane
conductivity.\cite{McKenzie98} Therefore the temperature
dependence of out-of-plane resistivity should follow the
temperature dependence of in-plane resistivity. Indeed, the
in-plane optical conductivity measurements on the Mott insulator
$\kappa \mbox{-(BEDT-TTF)}_2 \mbox{Cu[N(CN)}_2\mbox{]Cl}$, also
show that the Mott system becomes more metallic in a presence of
disorder. These measurements show the transfer of the spectral
weight to low frequency region as the irradiation time increases,
followed by the collapse of the Mott gap.\cite{Sasaki2007,
Sasaki2008}

\begin{figure}[t]
\begin{center}
\includegraphics[  width=2.5 in,
keepaspectratio]{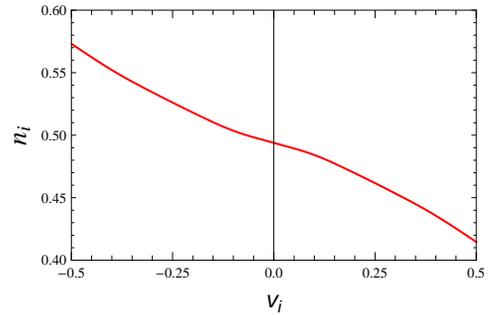}
\caption{(Color online) Local occupation number per spin as a
function of site disorder $v_i$ for $T=0.01, U=2.1$, and $W=1$.}
\label{occupation}%
\end{center}
\end{figure}

We emphasize that our model, as opposed to the physical mechanism
proposed in Ref.~\onlinecite{Sasaki2008}, does not assume an
introduction of new charge carriers since the total number of
carriers per site remains equal to one. The local occupation
number, however, depends on the random site potential, and we can
say that the system is effectively locally doped.\cite{HeidarianPRL2004}
The occupation number, for a given spin orientation, as a function of random site
potential is shown on Fig.~\ref{occupation}. It is interesting to
note that the local occupation number, $n(v_i)$, deviates much
less from its average value than it would be the case in the
absence of interaction. This is a consequence of very strong
disorder screening of site-diagonal disorder on the metallic side
of the Mott transition.\cite{Tanaskovic2003} Therefore, the
resistivity curves on Fig.~\ref{comparison}(a) cross at very low
temperatures and our current model cannot explain the intersection
of curves in Fig.~\ref{comparison}(b) which happens at much higher
temperature. The dramatic reduction of the elastic scattering is
also demonstrated in Ref.~\onlinecite{Aguiar2004}, which shows
that the inelastic scattering dominates in the incoherent regime.
We stress that we do not assume Matthiessen's rule. This is a
salient feature of DMFT, which can operate in a regime where
conventional approaches to the electron transport fail.

\section{Conclusions}

In summary, we have examined the influence of random potential on
the optical and DC conductivity for half-filled Hubbard model in a
vicinity of the Mott transition.  Our results show, in agreement
with the experiments on $\kappa$-organics, that the disorder can
make the system effectively more metallic. The disorder increases
the bandwidth which increases $U_c$ and weakens the correlation
effects, moves the system away from the Mott transition and leads
to a decrease in the scattering rate and resistivity. We emphasize
that the randomness in our model does not change global doping, as
the system remains on average half-filled, but the number of
charge carriers locally deviates from the average value.
Therefore, global carrier doping of a Mott insulator due to
irradiation defects, proposed in Ref.~\onlinecite{Sasaki2008}, is
not necessary to make the system more metallic. We also find that
the maximal possible value of metallic resistivity remains more
than an order of magnitude larger than the MIR limit even in a
presence of moderate disorder. As in the clean case, the violation
of the MIR limit is driven by a large scattering rate due to the
electron-electron scattering, and Drude-like peak in the optical
conductivity persists even at temperatures when the resistivity is
well beyond the MIR limit.

\acknowledgements The authors acknowledge fruitful discussions
with D.N.~Basov and N.E.~Hussey. This work was supported by the
Serbian Ministry of Science and Technological Development under
Project No.~OI 141035 (M.R. and D.T.), the NSF under Grant
No.~DMR-0542026 (V.D.) and the NSF Grant No.~DMR-0746395 (K.H).
M.R.~acknowledges support from the FP6 Center of Excellence grant
CX-CMCS and D.T.~support from the NATO Science for Peace and
Security Programme Reintegration Grant No. EAP.RIG.983235.
Numerical results were obtained on the AEGIS e-Infrastructure,
supported in part by FP7 projects EGEE-III and SEE-GRID-SCI.

\bibliographystyle{apsrev}
\bibliography{Radonjic_Incoherent_resub}

\begin{thebibliography}{32}
\expandafter\ifx\csname natexlab\endcsname\relax\def\natexlab#1{#1}\fi
\expandafter\ifx\csname bibnamefont\endcsname\relax
  \def\bibnamefont#1{#1}\fi
\expandafter\ifx\csname bibfnamefont\endcsname\relax
  \def\bibfnamefont#1{#1}\fi
\expandafter\ifx\csname citenamefont\endcsname\relax
  \def\citenamefont#1{#1}\fi
\expandafter\ifx\csname url\endcsname\relax
  \def\url#1{\texttt{#1}}\fi
\expandafter\ifx\csname urlprefix\endcsname\relax\def\urlprefix{URL }\fi
\providecommand{\bibinfo}[2]{#2}
\providecommand{\eprint}[2][]{\url{#2}}

\bibitem[{\citenamefont{Miranda and Dobrosavljevi\'{c}}(2005)}]{Miranda2005}
\bibinfo{author}{\bibfnamefont{E.}~\bibnamefont{Miranda}} \bibnamefont{and}
  \bibinfo{author}{\bibfnamefont{V.}~\bibnamefont{Dobrosavljevi\'{c}}},
  \bibinfo{journal}{Rep. Prog. Phys.} \textbf{\bibinfo{volume}{68}},
  \bibinfo{pages}{2337} (\bibinfo{year}{2005}).

\bibitem[{\citenamefont{Qazilbash et~al.}(2007)\citenamefont{Qazilbash, Brehm,
  Chae, Ho, Andreev, Kim, Yun, Balatsky, Maple, Keilmann et~al.}}]{Basov2007}
\bibinfo{author}{\bibfnamefont{M.~M.} \bibnamefont{Qazilbash}},
  \bibinfo{author}{\bibfnamefont{M.}~\bibnamefont{Brehm}},
  \bibinfo{author}{\bibfnamefont{B.-G.} \bibnamefont{Chae}},
  \bibinfo{author}{\bibfnamefont{P.-C.} \bibnamefont{Ho}},
  \bibinfo{author}{\bibfnamefont{G.~O.} \bibnamefont{Andreev}},
  \bibinfo{author}{\bibfnamefont{B.-J.} \bibnamefont{Kim}},
  \bibinfo{author}{\bibfnamefont{S.~J.} \bibnamefont{Yun}},
  \bibinfo{author}{\bibfnamefont{A.~V.} \bibnamefont{Balatsky}},
  \bibinfo{author}{\bibfnamefont{M.~B.} \bibnamefont{Maple}},
  \bibinfo{author}{\bibfnamefont{F.}~\bibnamefont{Keilmann}},
  \bibnamefont{et~al.}, \bibinfo{journal}{Science}
  \textbf{\bibinfo{volume}{318}}, \bibinfo{pages}{1750} (\bibinfo{year}{2007}).

\bibitem[{\citenamefont{Kohsaka et~al.}(2007)\citenamefont{Kohsaka, Taylor,
  Fujita, Schmidt, Lupien, Hanaguri, Azuma, Takano, Eisaki, Takagi
  et~al.}}]{Davis2007}
\bibinfo{author}{\bibfnamefont{Y.}~\bibnamefont{Kohsaka}},
  \bibinfo{author}{\bibfnamefont{C.}~\bibnamefont{Taylor}},
  \bibinfo{author}{\bibfnamefont{K.}~\bibnamefont{Fujita}},
  \bibinfo{author}{\bibfnamefont{A.}~\bibnamefont{Schmidt}},
  \bibinfo{author}{\bibfnamefont{C.}~\bibnamefont{Lupien}},
  \bibinfo{author}{\bibfnamefont{T.}~\bibnamefont{Hanaguri}},
  \bibinfo{author}{\bibfnamefont{M.}~\bibnamefont{Azuma}},
  \bibinfo{author}{\bibfnamefont{M.}~\bibnamefont{Takano}},
  \bibinfo{author}{\bibfnamefont{H.}~\bibnamefont{Eisaki}},
  \bibinfo{author}{\bibfnamefont{H.}~\bibnamefont{Takagi}},
  \bibnamefont{et~al.}, \bibinfo{journal}{Science}
  \textbf{\bibinfo{volume}{315}}, \bibinfo{pages}{1380} (\bibinfo{year}{2007}).

\bibitem[{\citenamefont{Analytis et~al.}(2006)\citenamefont{Analytis, Ardavan,
  Blundell, Owen, Garman, Jeynes, and Powell}}]{Analytis2006}
\bibinfo{author}{\bibfnamefont{J.~G.} \bibnamefont{Analytis}},
  \bibinfo{author}{\bibfnamefont{A.}~\bibnamefont{Ardavan}},
  \bibinfo{author}{\bibfnamefont{S.~J.} \bibnamefont{Blundell}},
  \bibinfo{author}{\bibfnamefont{R.~L.} \bibnamefont{Owen}},
  \bibinfo{author}{\bibfnamefont{E.~F.} \bibnamefont{Garman}},
  \bibinfo{author}{\bibfnamefont{C.}~\bibnamefont{Jeynes}}, \bibnamefont{and}
  \bibinfo{author}{\bibfnamefont{B.~J.} \bibnamefont{Powell}},
  \bibinfo{journal}{Phys. Rev. Lett.} \textbf{\bibinfo{volume}{96}},
  \bibinfo{pages}{177002} (\bibinfo{year}{2006}).

\bibitem[{\citenamefont{Sasaki et~al.}(2007)\citenamefont{Sasaki, Oizumi,
  Yoneyama, Kobayashi, and Toyota}}]{Sasaki2007}
\bibinfo{author}{\bibfnamefont{T.}~\bibnamefont{Sasaki}},
  \bibinfo{author}{\bibfnamefont{H.}~\bibnamefont{Oizumi}},
  \bibinfo{author}{\bibfnamefont{N.}~\bibnamefont{Yoneyama}},
  \bibinfo{author}{\bibfnamefont{N.}~\bibnamefont{Kobayashi}},
  \bibnamefont{and} \bibinfo{author}{\bibfnamefont{N.}~\bibnamefont{Toyota}},
  \bibinfo{journal}{J. Phys. Soc. Jpn.} \textbf{\bibinfo{volume}{76}},
  \bibinfo{pages}{123701} (\bibinfo{year}{2007}).

\bibitem[{\citenamefont{Sasaki et~al.}(2008)\citenamefont{Sasaki, Yoneyama,
  Nakamura, Kobayashi, Ikemoto, Moriwaki, and Kimura}}]{Sasaki2008}
\bibinfo{author}{\bibfnamefont{T.}~\bibnamefont{Sasaki}},
  \bibinfo{author}{\bibfnamefont{N.}~\bibnamefont{Yoneyama}},
  \bibinfo{author}{\bibfnamefont{Y.}~\bibnamefont{Nakamura}},
  \bibinfo{author}{\bibfnamefont{N.}~\bibnamefont{Kobayashi}},
  \bibinfo{author}{\bibfnamefont{Y.}~\bibnamefont{Ikemoto}},
  \bibinfo{author}{\bibfnamefont{T.}~\bibnamefont{Moriwaki}}, \bibnamefont{and}
  \bibinfo{author}{\bibfnamefont{H.}~\bibnamefont{Kimura}},
  \bibinfo{journal}{Phys. Rev. Lett.} \textbf{\bibinfo{volume}{101}},
  \bibinfo{pages}{206403} (\bibinfo{year}{2008}).

\bibitem[{\citenamefont{Powell and McKenzie}(2006)}]{Powell2006}
\bibinfo{author}{\bibfnamefont{B.~J.} \bibnamefont{Powell}} \bibnamefont{and}
  \bibinfo{author}{\bibfnamefont{R.~H.} \bibnamefont{McKenzie}},
  \bibinfo{journal}{J. Phys.: Condens. Matter} \textbf{\bibinfo{volume}{18}},
  \bibinfo{pages}{R827} (\bibinfo{year}{2006}).

\bibitem[{\citenamefont{Gunnarsson et~al.}(2003)\citenamefont{Gunnarsson,
  Calandra, and Han}}]{Gunnarsson2003}
\bibinfo{author}{\bibfnamefont{O.}~\bibnamefont{Gunnarsson}},
  \bibinfo{author}{\bibfnamefont{M.}~\bibnamefont{Calandra}}, \bibnamefont{and}
  \bibinfo{author}{\bibfnamefont{J.~E.} \bibnamefont{Han}},
  \bibinfo{journal}{Rev. Mod. Phys.} \textbf{\bibinfo{volume}{75}},
  \bibinfo{pages}{1085} (\bibinfo{year}{2003}).

\bibitem[{\citenamefont{Calandra and Gunnarsson}(2001)}]{CalandraPRL2001}
\bibinfo{author}{\bibfnamefont{M.}~\bibnamefont{Calandra}} \bibnamefont{and}
  \bibinfo{author}{\bibfnamefont{O.}~\bibnamefont{Gunnarsson}},
  \bibinfo{journal}{Phys. Rev. Lett.} \textbf{\bibinfo{volume}{87}},
  \bibinfo{pages}{266601} (\bibinfo{year}{2001}).

\bibitem[{\citenamefont{Calandra and Gunnarsson}(2002)}]{CalandraPRB2002}
\bibinfo{author}{\bibfnamefont{M.}~\bibnamefont{Calandra}} \bibnamefont{and}
  \bibinfo{author}{\bibfnamefont{O.}~\bibnamefont{Gunnarsson}},
  \bibinfo{journal}{Phys. Rev. B} \textbf{\bibinfo{volume}{66}},
  \bibinfo{pages}{205105} (\bibinfo{year}{2002}).

\bibitem[{\citenamefont{Allen}(2002)}]{Allen2002}
\bibinfo{author}{\bibfnamefont{M.~B.} \bibnamefont{Allen}},
  \bibinfo{journal}{Physica B} \textbf{\bibinfo{volume}{318}},
  \bibinfo{pages}{24} (\bibinfo{year}{2002}).

\bibitem[{\citenamefont{Hussey et~al.}(2004)\citenamefont{Hussey, Takenaka, and
  Takagi}}]{Hussey2004}
\bibinfo{author}{\bibfnamefont{N.~E.} \bibnamefont{Hussey}},
  \bibinfo{author}{\bibfnamefont{K.}~\bibnamefont{Takenaka}}, \bibnamefont{and}
  \bibinfo{author}{\bibfnamefont{H.}~\bibnamefont{Takagi}},
  \bibinfo{journal}{Philos. Mag.} \textbf{\bibinfo{volume}{84}},
  \bibinfo{pages}{2847} (\bibinfo{year}{2004}).

\bibitem[{\citenamefont{Merino and McKenzie}(2000)}]{Merino2000}
\bibinfo{author}{\bibfnamefont{J.}~\bibnamefont{Merino}} \bibnamefont{and}
  \bibinfo{author}{\bibfnamefont{R.~H.} \bibnamefont{McKenzie}},
  \bibinfo{journal}{Phys. Rev. B} \textbf{\bibinfo{volume}{61}},
  \bibinfo{pages}{7996} (\bibinfo{year}{2000}).

\bibitem[{\citenamefont{Limelette et~al.}(2003)\citenamefont{Limelette,
  Wzietek, Florens, Georges, Costi, Pasquier, J\'{e}rome, M\'{e}zi\`{e}re, and
  Batail}}]{Limelette2003}
\bibinfo{author}{\bibfnamefont{P.}~\bibnamefont{Limelette}},
  \bibinfo{author}{\bibfnamefont{P.}~\bibnamefont{Wzietek}},
  \bibinfo{author}{\bibfnamefont{S.}~\bibnamefont{Florens}},
  \bibinfo{author}{\bibfnamefont{A.}~\bibnamefont{Georges}},
  \bibinfo{author}{\bibfnamefont{T.~A.} \bibnamefont{Costi}},
  \bibinfo{author}{\bibfnamefont{C.}~\bibnamefont{Pasquier}},
  \bibinfo{author}{\bibfnamefont{D.}~\bibnamefont{J\'{e}rome}},
  \bibinfo{author}{\bibfnamefont{C.}~\bibnamefont{M\'{e}zi\`{e}re}},
  \bibnamefont{and} \bibinfo{author}{\bibfnamefont{P.}~\bibnamefont{Batail}},
  \bibinfo{journal}{Phys. Rev. Lett.} \textbf{\bibinfo{volume}{91}},
  \bibinfo{pages}{016401} (\bibinfo{year}{2003}).

\bibitem[{\citenamefont{Merino et~al.}(2008)\citenamefont{Merino, Dumm,
  Drichko, Dressel, and McKenzie}}]{Merino2008}
\bibinfo{author}{\bibfnamefont{J.}~\bibnamefont{Merino}},
  \bibinfo{author}{\bibfnamefont{M.}~\bibnamefont{Dumm}},
  \bibinfo{author}{\bibfnamefont{N.}~\bibnamefont{Drichko}},
  \bibinfo{author}{\bibfnamefont{M.}~\bibnamefont{Dressel}}, \bibnamefont{and}
  \bibinfo{author}{\bibfnamefont{R.~H.} \bibnamefont{McKenzie}},
  \bibinfo{journal}{Phys. Rev. Lett.} \textbf{\bibinfo{volume}{100}},
  \bibinfo{pages}{086404} (\bibinfo{year}{2008}).

\bibitem[{\citenamefont{Laad et~al.}(2001)\citenamefont{Laad, Craco, and M{\"
  u}ller-Hartmann}}]{Laad2001}
\bibinfo{author}{\bibfnamefont{M.~S.} \bibnamefont{Laad}},
  \bibinfo{author}{\bibfnamefont{L.}~\bibnamefont{Craco}}, \bibnamefont{and}
  \bibinfo{author}{\bibfnamefont{E.}~\bibnamefont{M{\" u}ller-Hartmann}},
  \bibinfo{journal}{Phys. Rev. B} \textbf{\bibinfo{volume}{64}},
  \bibinfo{pages}{195114} (\bibinfo{year}{2001}).

\bibitem[{\citenamefont{Dobrosavljevi\'{c} and
  Kotliar}(1997)}]{Dobrosavljevic1997}
\bibinfo{author}{\bibfnamefont{V.}~\bibnamefont{Dobrosavljevi\'{c}}}
  \bibnamefont{and} \bibinfo{author}{\bibfnamefont{G.}~\bibnamefont{Kotliar}},
  \bibinfo{journal}{Phys. Rev. Lett.} \textbf{\bibinfo{volume}{78}},
  \bibinfo{pages}{3943} (\bibinfo{year}{1997}).

\bibitem[{\citenamefont{Dobrosavljevi\'{c}
  et~al.}(2003)\citenamefont{Dobrosavljevi\'{c}, Pastor, and
  Nikoli\'{c}}}]{Dobrosavljevic2003}
\bibinfo{author}{\bibfnamefont{V.}~\bibnamefont{Dobrosavljevi\'{c}}},
  \bibinfo{author}{\bibfnamefont{A.~A.} \bibnamefont{Pastor}},
  \bibnamefont{and} \bibinfo{author}{\bibfnamefont{B.~K.}
  \bibnamefont{Nikoli\'{c}}}, \bibinfo{journal}{Europhys. Lett.}
  \textbf{\bibinfo{volume}{62}}, \bibinfo{pages}{76} (\bibinfo{year}{2003}).

\bibitem[{\citenamefont{Byczuk et~al.}(2005)\citenamefont{Byczuk, Hofstetter,
  and Vollhardt}}]{Byczuk2005}
\bibinfo{author}{\bibfnamefont{K.}~\bibnamefont{Byczuk}},
  \bibinfo{author}{\bibfnamefont{W.}~\bibnamefont{Hofstetter}},
  \bibnamefont{and}
  \bibinfo{author}{\bibfnamefont{D.}~\bibnamefont{Vollhardt}},
  \bibinfo{journal}{Phys. Rev. Lett.} \textbf{\bibinfo{volume}{94}},
  \bibinfo{pages}{056404} (\bibinfo{year}{2005}).

\bibitem[{\citenamefont{Economou}(2005)}]{Economou2005}
\bibinfo{author}{\bibfnamefont{E.~N.} \bibnamefont{Economou}},
  \emph{\bibinfo{title}{Green’s Functions in Quantum Physics}}
  (\bibinfo{publisher}{Springer}, \bibinfo{address}{Berlin},
  \bibinfo{year}{2005}), \bibinfo{edition}{3rd} ed.

\bibitem[{\citenamefont{Pruschke et~al.}(1993)\citenamefont{Pruschke, Cox, and
  Jarrell}}]{Pruschke1993}
\bibinfo{author}{\bibfnamefont{T.}~\bibnamefont{Pruschke}},
  \bibinfo{author}{\bibfnamefont{D.~L.} \bibnamefont{Cox}}, \bibnamefont{and}
  \bibinfo{author}{\bibfnamefont{M.}~\bibnamefont{Jarrell}},
  \bibinfo{journal}{Phys. Rev. B} \textbf{\bibinfo{volume}{47}},
  \bibinfo{pages}{3553} (\bibinfo{year}{1993}).

\bibitem[{\citenamefont{Haule et~al.}(2001)\citenamefont{Haule, Kirchner,
  Kroha, and W{\" o}lfle}}]{Haule2001}
\bibinfo{author}{\bibfnamefont{K.}~\bibnamefont{Haule}},
  \bibinfo{author}{\bibfnamefont{S.}~\bibnamefont{Kirchner}},
  \bibinfo{author}{\bibfnamefont{J.}~\bibnamefont{Kroha}}, \bibnamefont{and}
  \bibinfo{author}{\bibfnamefont{P.}~\bibnamefont{W{\" o}lfle}},
  \bibinfo{journal}{Phys. Rev. B} \textbf{\bibinfo{volume}{64}},
  \bibinfo{pages}{155111} (\bibinfo{year}{2001}).

\bibitem[{\citenamefont{Werner et~al.}(2006)\citenamefont{Werner, Comanac,
  de’ Medici, Troyer, and Millis}}]{Werner2006}
\bibinfo{author}{\bibfnamefont{P.}~\bibnamefont{Werner}},
  \bibinfo{author}{\bibfnamefont{A.}~\bibnamefont{Comanac}},
  \bibinfo{author}{\bibfnamefont{L.}~\bibnamefont{de’ Medici}},
  \bibinfo{author}{\bibfnamefont{M.}~\bibnamefont{Troyer}}, \bibnamefont{and}
  \bibinfo{author}{\bibfnamefont{A.~J.} \bibnamefont{Millis}},
  \bibinfo{journal}{Phys. Rev. Lett.} \textbf{\bibinfo{volume}{97}},
  \bibinfo{pages}{076405} (\bibinfo{year}{2006}).

\bibitem[{\citenamefont{Haule}(2007)}]{Haule2007}
\bibinfo{author}{\bibfnamefont{K.}~\bibnamefont{Haule}},
  \bibinfo{journal}{Phys. Rev. B} \textbf{\bibinfo{volume}{75}},
  \bibinfo{pages}{155113} (\bibinfo{year}{2007}).

\bibitem[{\citenamefont{Liebsch et~al.}(2009)\citenamefont{Liebsch, Ishida, and
  Merino}}]{Liebsch2009}
\bibinfo{author}{\bibfnamefont{A.}~\bibnamefont{Liebsch}},
  \bibinfo{author}{\bibfnamefont{H.}~\bibnamefont{Ishida}}, \bibnamefont{and}
  \bibinfo{author}{\bibfnamefont{J.}~\bibnamefont{Merino}},
  \bibinfo{journal}{Phys. Rev. B} \textbf{\bibinfo{volume}{79}},
  \bibinfo{pages}{195108} (\bibinfo{year}{2009}).

\bibitem[{\citenamefont{Georges et~al.}(1996)\citenamefont{Georges, Kotliar,
  Krauth, and Rozenberg}}]{Georges1996}
\bibinfo{author}{\bibfnamefont{A.}~\bibnamefont{Georges}},
  \bibinfo{author}{\bibfnamefont{G.}~\bibnamefont{Kotliar}},
  \bibinfo{author}{\bibfnamefont{W.}~\bibnamefont{Krauth}}, \bibnamefont{and}
  \bibinfo{author}{\bibfnamefont{M.~J.} \bibnamefont{Rozenberg}},
  \bibinfo{journal}{Rev. Mod. Phys.} \textbf{\bibinfo{volume}{68}},
  \bibinfo{pages}{13} (\bibinfo{year}{1996}).

\bibitem[{\citenamefont{Aguiar et~al.}(2005)\citenamefont{Aguiar,
  Dobrosavljevi\'{c}, Abrahams, and Kotliar}}]{Aguiar2005}
\bibinfo{author}{\bibfnamefont{M.~C.} \bibnamefont{Aguiar}},
  \bibinfo{author}{\bibfnamefont{V.}~\bibnamefont{Dobrosavljevi\'{c}}},
  \bibinfo{author}{\bibfnamefont{E.}~\bibnamefont{Abrahams}}, \bibnamefont{and}
  \bibinfo{author}{\bibfnamefont{G.}~\bibnamefont{Kotliar}},
  \bibinfo{journal}{Phys. Rev. B} \textbf{\bibinfo{volume}{71}},
  \bibinfo{pages}{205115} (\bibinfo{year}{2005}).

\bibitem[{\citenamefont{Qazilbash et~al.}(2006)\citenamefont{Qazilbash, Burch,
  Whisler, Shrekenhamer, Chae, Kim, and Basov}}]{Basov2006}
\bibinfo{author}{\bibfnamefont{M.~M.} \bibnamefont{Qazilbash}},
  \bibinfo{author}{\bibfnamefont{K.~S.} \bibnamefont{Burch}},
  \bibinfo{author}{\bibfnamefont{D.}~\bibnamefont{Whisler}},
  \bibinfo{author}{\bibfnamefont{D.}~\bibnamefont{Shrekenhamer}},
  \bibinfo{author}{\bibfnamefont{B.~G.} \bibnamefont{Chae}},
  \bibinfo{author}{\bibfnamefont{H.~T.} \bibnamefont{Kim}}, \bibnamefont{and}
  \bibinfo{author}{\bibfnamefont{D.~N.} \bibnamefont{Basov}},
  \bibinfo{journal}{Phys. Rev. B} \textbf{\bibinfo{volume}{74}},
  \bibinfo{pages}{205118} (\bibinfo{year}{2006}).

\bibitem[{\citenamefont{McKenzie and Moses}(1998)}]{McKenzie98}
\bibinfo{author}{\bibfnamefont{R.~H.} \bibnamefont{McKenzie}} \bibnamefont{and}
  \bibinfo{author}{\bibfnamefont{P.}~\bibnamefont{Moses}},
  \bibinfo{journal}{Phys. Rev. Lett.} \textbf{\bibinfo{volume}{81}},
  \bibinfo{pages}{4492} (\bibinfo{year}{1998}).

\bibitem[{\citenamefont{Heidarian and Trivedi}(2004)}]{HeidarianPRL2004}
\bibinfo{author}{\bibfnamefont{D.}~\bibnamefont{Heidarian}} \bibnamefont{and}
  \bibinfo{author}{\bibfnamefont{N.}~\bibnamefont{Trivedi}},
  \bibinfo{journal}{Phys. Rev. Lett.} \textbf{\bibinfo{volume}{93}},
  \bibinfo{pages}{126401} (\bibinfo{year}{2004}).

\bibitem[{\citenamefont{Tanaskovi{\'c}
  et~al.}(2003)\citenamefont{Tanaskovi{\'c}, Dobrosavljevi\'{c}, Abrahams, and
  Kotliar}}]{Tanaskovic2003}
\bibinfo{author}{\bibfnamefont{D.}~\bibnamefont{Tanaskovi{\'c}}},
  \bibinfo{author}{\bibfnamefont{V.}~\bibnamefont{Dobrosavljevi\'{c}}},
  \bibinfo{author}{\bibfnamefont{E.}~\bibnamefont{Abrahams}}, \bibnamefont{and}
  \bibinfo{author}{\bibfnamefont{G.}~\bibnamefont{Kotliar}},
  \bibinfo{journal}{Phys. Rev. Lett.} \textbf{\bibinfo{volume}{91}},
  \bibinfo{pages}{066603} (\bibinfo{year}{2003}).

\bibitem[{\citenamefont{Aguiar et~al.}(2004)\citenamefont{Aguiar, Miranda,
  Dobrosavljevi\'{c}, Abrahams, and Kotliar}}]{Aguiar2004}
\bibinfo{author}{\bibfnamefont{M.~C.} \bibnamefont{Aguiar}},
  \bibinfo{author}{\bibfnamefont{E.}~\bibnamefont{Miranda}},
  \bibinfo{author}{\bibfnamefont{V.}~\bibnamefont{Dobrosavljevi\'{c}}},
  \bibinfo{author}{\bibfnamefont{E.}~\bibnamefont{Abrahams}}, \bibnamefont{and}
  \bibinfo{author}{\bibfnamefont{G.}~\bibnamefont{Kotliar}},
  \bibinfo{journal}{Europhys. Lett.} \textbf{\bibinfo{volume}{67}},
  \bibinfo{pages}{226} (\bibinfo{year}{2004}).

\end{thebibliography}

\end{document}